\documentclass[11pt,a4paper]{article}
\pdfoutput=1 
\usepackage{fullpage}
\usepackage{graphicx}%
\usepackage{multirow}%
\usepackage{amsmath,amssymb,amsfonts}%
\usepackage{amsthm}%
\usepackage{mathrsfs}%
\usepackage[title]{appendix}%
\usepackage{xcolor}%
\usepackage{textcomp}%
\usepackage{manyfoot}%
\usepackage{booktabs}%
\usepackage{algorithmicx}%
\usepackage{algpseudocode}%
\usepackage{listings}%
\usepackage[numbers]{natbib}
\usepackage{mathtools}
\usepackage{hyperref}
\usepackage{cleveref}
\usepackage[linesnumbered,ruled,vlined]{algorithm2e}
\usepackage{multicol}
\usepackage{enumerate}
\usepackage{makecell}
\usepackage{bm}
\usepackage{subcaption}
\usepackage[inline]{enumitem}
\newlist{properties}{enumerate}{1}
\setlist[properties]{label={(\alph*)}}
\crefname{propertiesi}{Property}{Properties}
\newlist{iproperties}{enumerate*}{1}
\setlist[iproperties]{label={(\alph*)}}
\crefname{ipropertiesi}{Property}{Properties}

\usepackage{thmtools} 
\usepackage{thm-restate}
\usepackage{tcolorbox}
\usepackage{xspace}
\usepackage[bottom]{footmisc}

\usepackage{doi}

\renewcommand{\paragraph}[1]{\medskip\noindent\textbf{#1.\;}}

\renewcommand{\le}{\leqslant}
\renewcommand{\leq}{\leqslant}
\renewcommand{\ge}{\geqslant}
\renewcommand{\geq}{\geqslant}

\renewcommand{\hat}{\widehat}

\newcommand{\sv}{\ensuremath{\vec s}}
\newcommand{\tran}{\intercal}

\newcommand{\voters}{\ensuremath{N}}
\newcommand{\cands}{\ensuremath{A}}
\newcommand{\alts}{\ensuremath{A}}
\newcommand{\cd}{\ensuremath{a}}

\newcommand{\opt}{\cd^*}

\newcommand{\sw}{\ensuremath{\mathsf{sw}}}

\DeclarePairedDelimiter{\set}{\{}{\}}

\DeclareMathOperator{\E}{\mathbb{E}}

\newcommand{\dist}{\ensuremath{\mathsf{dist}}}

\newcommand{\utp}{\vec u}
\newcommand{\prefp}{\vec \pref}

\DeclareMathOperator*{\argmax}{\arg\max}
\DeclareMathOperator{\argmin}{\arg\min}

\usepackage{bbm}

\newcommand{\rank}{\rk}

\newcommand{\pref}{\ensuremath{\mathsf{\sigma}}}

\newcommand{\ind}{\ensuremath{\mathbb{I}}}

\newcommand{\dich}{{\mathbbm{1}}}
\newcommand{\calC}{\ensuremath{\mathcal{C}}}

\newcommand{\vs}{\ensuremath{\vec{s}}}

\newcommand{\prs}{\mathsf{P}_{\sv}}
\newcommand{\rk}{\mathsf{rank}}
\newcommand{\btl}{BCHLPS}

\newcommand{\PP}{$\mathcal{P}$\xspace}
\newcommand{\QQ}{$\mathcal{Q}$\xspace}
\newtheorem{theorem}{Theorem}

\newtheorem{lemma}{Lemma}

\newtheorem{definition}[theorem]{Definition}

\theoremstyle{remark}

\crefname{algocf}{algorithm}{algorithms}

\title{Computational Aspects of Distortion}

\author{
\begin{tabular}{c c}
& \\ \textbf{Soroush Ebadian} & \textbf{Aris Filos-Ratsikas}\\
\small{University of Toronto, Canada} & \small{Imperial College London, UK} \\
\href{mailto:soroush@cs.toronto.edu}{\small{\texttt{soroush@cs.toronto.edu}}} & \href{mailto:aris.filos-ratsikas@imperial.ac.uk}{\small{\texttt{aris.filos-ratsikas@imperial.ac.uk}}}\\
& \\
\textbf{Mohamad Latifian} & \textbf{Nisarg Shah}\\
\small{University of Edinburgh, UK} & \small{University of Toronto, Canada} \\
\href{mailto:mohamad.latifian@ed.ac.uk}{\small{\texttt{mohamad.latifian@ed.ac.uk}}} & \href{mailto:nisarg@cs.toronto.edu}{\small{\texttt{nisarg@cs.toronto.edu}}}
\end{tabular}}

\date{}

\begin{document}
\maketitle

\abstract{
The distortion framework in social choice theory allows quantifying the efficiency of (randomized) selection of an alternative based on the preferences of a set of agents. We make two fundamental contributions to this framework. 
First, we develop a linear-programming-based algorithm for computing the optimal randomized decision on a given instance, which is simpler and faster than the state-of-the-art solutions. For practitioners who may prefer to deploy a classical decision-making rule over the aforementioned optimal rule, we develop an algorithm based on non-convex quadratic programming for computing the \emph{exact} distortion of any (and the best) randomized positional scoring rule. For a small number of alternatives, we find that the exact distortion bounds are significantly better than the asymptotic upper bounds established in prior literature and lead to different recommendations on which rules to use.  
These results rely on a novel characterization of the instances yielding the worst distortion, which may be of independent interest. 
}

\section{Introduction}

The area of computational social choice aims to find good ways of aggregating the preferences of a set of people, or \emph{voters}, into a desirable collective decision. 
The notion of distortion in social choice was introduced by \citet{PR06} in 2006 to measure the inefficiency of voting rules that only have access to limited information about the preferences of the agents. Over the past two decades, this literature has given rise to a plethora of interesting results about the distortion of rules in several different domains of interest, including single-winner and committee voting \cite{BCHL+15,caragiannis2017subset,borodin2022distortion,bhaskar2018truthful,amanatidis2021peeking,mandal2019efficient,mandal2020optimal,anshelevich2018approximating}, participatory budgeting \cite{benade2021preference,goyal2023low}, and matching \cite{filos2014social,amanatidis2022few,anshelevich2021ordinal,caragiannis2022truthful,anari2023distortion}.

One of the most fundamental settings that have been studied under this framework is that of implicit utilitarian voting, in which the preferences of the agents are expressed by normalized but otherwise unrestricted cardinal values, and the voting rules only have access to the ordinal preference rankings induced by those values. For this setting, some of the very first works in the area already identified the limitations of voting rules, even those that employ randomization in their decisions, by providing a general lower bound of $\Omega(\sqrt{m})$ on the best achievable distortion \cite{BCHL+15}, where $m$ is the number of alternatives. Far more recently, \citet{EKPS22} designed a voting rule that always achieves a distortion of $O(\sqrt{m})$, thus settling the asymptotic efficiency of general voting rules. Even more recently, \citet{ebadian2023explainable} studied the distortion of a class of very natural randomized voting rules, called randomized positional scoring rules (RPSRs), which select an outcome proportionally to its score, according to some predefined scoring vector. Rules in this class display several attractive properties (such as anonymity, neutrality, and strategyproofness \citep{barbera1978nice,gibbard1977manipulation}) and in \cite{ebadian2023explainable} are deemed \emph{explainable}, as they use randomization in a rather straightforward manner. The best achievable asymptotic distortion by RPSRs has been shown to be $\Theta (\sqrt{m \log m})$ \cite{BCHL+15, bhaskar2018truthful}, and the asymptotic distortion bounds for several interesting RPSRs are known~\cite{ebadian2023explainable}.

From the preceding paragraphs, it is evident that the \emph{worst-case, asymptotic} distortion of voting rules for implicit utilitarian voting is pretty well understood. Still, many of the preference vectors that we encounter in applications exhibit a certain structure, which does not necessarily allow for the pathological examples that establish the worst-case lower bounds to arise, raising the following question:\smallskip

\noindent\emph{
    1. Given a vector of agents' preferences, how can we compute the best possible distortion on this vector? How can we find a voting rule that achieves it?
}\smallskip

\noindent Additionally, even in a worst-case regime, in many cases we should not be satisfied with merely asymptotic bounds, as the hidden constants, even when small, can have a significant impact on the calculated distortion bounds. This latter fact is particularly pronounced in instances with a small number of possible outcomes, which are often encountered in practice. Indeed, elections for political leaders often exhibit a small number of alternatives, or selecting a candidate for an academic position usually boils down to a choice between a few shortlisted options. Especially for those cases, we would like to be able to compute the \emph{exact} distortion bound of a given voting rule, captured in the following question: \smallskip

\noindent\emph{
    2. Given a voting rule, how can we find its \emph{exact} worst-case distortion?
}

\subsection{Our contributions}

In this paper, we provide algorithms that answer Questions 1 and 2 above, which are based on continuous optimization techniques. In particular, we provide the following contributions for the distortion of voting rules in the \emph{unit-sum normalized} utilitarian setting. \medskip

\subsubsection*{For question 1} We design a polynomial-time algorithm which, for any vector of agents' (ordinal) preferences, i.e., for any \emph{preference profile} with $n$ agents and $m$ alternatives, constructs a linear program with $O(nm)$ variables and $O(nm)$ constraints which computes the best possible distortion, and the voting rule that achieves this distortion. Our algorithm is conceptually and computationally simpler than the best-known solution due to \citep{BCHL+15}, which also employs a linear program with $O(nm^2)$ variables and constraints. For the first part of Question 1 in isolation, i.e., merely identifying the best possible distortion of any voting rule given a preference profile (but not the rule itself), we provide a much faster algorithm, which runs in time $O(nm\log(nm))$. We complement our theoretical results with experiments which demonstrate the far superior running time of our algorithms on both real and synthetic data. \medskip

\subsubsection*{For question 2} We formulate the problem of finding the \emph{exact} distortion of a given RPSR as a quadratic program. This quadratic program is non-convex, and thus our algorithm does not run in polynomial time. Still, it is fast enough to comfortably compute the exact distortion of a given RPSR for up to $5$ candidates, which captures a plethora of social choice scenarios of interest. Our algorithm is the first to solve the exact distortion computation problem, but in the next section, we do highlight its computational advantages over previous techniques that could conceivably be applied to our domain. Using this quadratic program, we compute the exact distortion of several well-known RPSRs, as well as the best possible distortion achieved by any rule in this class. For the latter part, we employ an iterative technique that searches through the (continuous) domain of RPSRs, aided by distortion lower bounds for rules in this class obtained as solutions to linear programs, making use of classic characterization results from the literature \cite{gibbard1977manipulation,barbera1978nice}. \medskip

\noindent All of our results are enabled by a novel characterization of the structure of the worst-case instances for any (ordinal) voting rule, by means of what we refer to as \emph{dichotomous utilities}, which could be of independent interest. 

\subsection{Related work and discussion}\label{sec:related}

We discuss how our results for Questions 1 and 2 improve upon the state-of-the-art results in the literature. For more works on the general topic of distortion in computational social choice theory, we defer the reader to the survey of \citet{anshelevich2021distortion}.

For finding the preference profile-optimal rule and its distortion in Question 1, \citet{BCHL+15} provided a polynomial-time algorithm based on linear programming. At a high level, their linear program finds the smallest possible distortion value subject to a set of constructed inequalities that ensure that this distortion is feasible for some voting rule. In turn, these inequalities are obtained from the dual of a different linear program that checks the feasibility of a given distortion; the use of duality here is crucial, to avoid their resulting optimization program having quadratic constraints. \citet{BCHL+15}, as also evidenced by the phrasing of Theorem 3.4 in their work, primarily presented their algorithm as a proof of polynomial-time solvability of the problem. That said, merely establishing polynomial-time solvability can be achieved by a simpler linear program, one with infinitely many constraints, coupled with an appropriate separation oracle so that it can be solved by the ellipsoid method; see \Cref{sec:optimal-distribution} for more details. Still, even the algorithm of \cite{BCHL+15} turns out to not be fast enough for several applications, e.g., for computing the preference profile-optimal rule on a large set of inputs, to be used as a performance benchmark against other voting rules in experiments. Our algorithms are faster in theory and in practice, making them much more appealing for such applications. The main technical contribution that allows us to devise these faster algorithms is our novel characterization of dichotomous utilities as worst-case utility profiles. 

Finally, we remark that one can interpret our algorithm as a general social choice rule with a best-possible distortion of $O(\sqrt m)$. Compared to the rule of \citet{EKPS22}, ours seems conceptually simpler, as it is based on linear programs. In contrast, the rule in \citep{EKPS22} uses the concept of \emph{stable lotteries}, which are computed by applying a multiplicative-weights update algorithm to approach the value of a certain zero-sum game, see \cite{jiang2020approximately} for more details. 

Moving on to the exact distortion of RPSRs in Question 2, the literature had not really provided any methods for this task prior to our work. The most related approach is due to \citet{filos2014truthful}, who studied the exact distortion of RPSRs for three candidates, among other voting rules, but crucially, for a different normalization of the utilities called \emph{unit-range}.\footnote{In unit-range, the values of each agent for the alternatives lie in $[0,1]$ with the maximum value being $1$ and the minimum value being $0$. In unit-sum, the values of each agent for the alternatives sum to $1$. The unit-sum normalization is the most widely used in the related literature, see \citep{anshelevich2021distortion,aziz2020justifications}.} Given our structural characterization of the worst-case utility profiles, their approach is in principle applicable for unit-sum as well, but it results in non-convex quadratic programs with at least $18$ variables and constraints, even for $m=3$. Our program, which is different from the one of \citep{filos2014truthful}, has far fewer constraints and can thus easily handle instances with $m=5$ alternatives; we provide more details about the comparison with \citep{filos2014truthful} in \Cref{app:filos-ratsikas-miltersen}.


To find the best RPSR and its distortion, we repeatedly apply our quadratic program on a set of candidate RPSRs. To guide our search, we follow another idea of \citep{filos2014truthful}, and obtain those candidates as solutions to a sequence of zero-sum games, which can easily be solved via linear programming. The zero-sum game formulation is enabled by well-known characterizations of RPSRs due to \citet{gibbard1977manipulation} and \citet{barbera1978nice} (see also \citep{filos2014truthful}), and can also be seen as an application of \citeauthor{yao1977probabilistic}'s minimax principle \citep{yao1977probabilistic}.


\section{Preliminaries} 
Let $[t] := \set{1, 2, \ldots, t}$ for $t \in \mathbb N$. For a set $S$, let $\Delta(S)$ be the set of probability distributions over $S$. 

\vspace{0.25em}
\noindent\emph{Utilitarian voting.}
Let $\voters$ be a set of $n$ agents and $\alts$ be a set of $m$ alternatives. We assume that each agent $i \in \voters$ has a utility function $u_i : \alts \to \mathbb{R}_{\ge 0}$ over the alternatives. Following the literature, we adopt the unit-sum assumption: $\sum_{a \in \alts} u_i(a) = 1$ for each $i \in \voters$~\cite{aziz2020justifications}. These utility functions collectively form the utility profile $\utp$. With slight abuse of notation, we use $u_i(p) := \E_{a \sim p} u_i(a)$ to denote the (expected) utility of agent $i$ under distribution $p \in \Delta(\alts)$. Given a utility profile $\utp$, the social welfare of an alternative $a \in \alts$ is $\sw(a,\utp) := \sum_{i \in \voters} u_i(a)$ and that of a distribution $p \in \Delta(\alts)$ is $\sw(p, \utp) := \sum_{i \in \voters} u_i(p)$. We say that an alternative $a^*$ is optimal if it has the maximum social welfare, i.e., $\opt(\utp) \in \argmax_{a \in \alts} \sw(a, \utp)$. We might drop $\utp$ when it is clear from the context.

\subsubsection*{Elicitation and aggregation}
Based on their underlying utilities, agents submit their votes in the form of a ranking over the alternatives. Let $\pref_i : [m] \to \alts$ denote the ranking of agent $i \in \voters$ over the alternatives. We use $a \succ_i b$ to show that agent $i$ prefers alternative $a$ to alternative $b$, and $\rk_i(a) := \pref^{-1}_i(a)$ to show the rank of alternative $a$ in agent $i$'s ranking; thus $a \succ_i b \Leftrightarrow \rk_i(a) < \rk_i(b)$. These rankings collectively form a preference profile $\prefp$. 
We say that $\pref_i$ is consistent with utility function $u_i$ if, for any pair of alternatives $a, b \in \alts$, $a \succ_i b$ implies $u_i(a) \geq u_i(b)$; let $\calC(\pref_i)$ denote the set of utility functions consistent with $\pref_i$.  We say that utility profile $\utp$ is consistent with $\prefp$ if $u_i \in \calC(\pref_i)$ for each agent $i \in \voters$; let $\calC(\prefp)$ denote the set of utility profiles consistent with $\prefp$.

\subsubsection*{Voting rules} 
A (randomized) voting rule $f$ takes a preference profile $\prefp$ as input and outputs a distribution $f(\prefp) \in \Delta(\alts)$ over the alternatives. If $f(\prefp)$ always has singleton support, we say $f$ is \emph{deterministic} and, with slight abuse of notation, use $f(\prefp)$ to denote the alternative in the support. 

\subsubsection*{Distortion} The distortion of a distribution $p \in \Delta(\alts)$ on a utility profile $\utp$ is its social welfare approximation,
$\dist(p,\utp) := \frac{\max_{a \in \alts} \sw(a,\utp)}{\sw(p,\utp)}$. 
Its distortion on a preference profile $\prefp$ is its worst-case distortion on any utility profile $\utp \in \calC(\prefp)$: $\dist(p,\prefp) := \sup_{\utp \in \calC(\prefp)} \dist(p,\utp)$; note that this supremum is attained at some $\utp \in \calC(\prefp)$ due to a continuous function being optimized over a compact domain and, thus, can be replaced by a maximum. The distortion of a voting rule $f$ is the worst-case distortion of its output over all preference profiles: $\dist(f) = \sup_{\prefp} \dist(f(\prefp),\prefp)$, where the worst case is taken over all preference profiles with $m$ alternatives (and any number of agents). The \emph{instance-optimal rule} $f^*$ is the rule that, on each preference profile $\prefp$, outputs the distribution with the smallest distortion: $f^*(\prefp) \in \argmin_{p \in \Delta(\alts)} \dist(p,\prefp)$. 

\section{Computing the optimal distribution}\label{sec:optimal-distribution}

In this section, we focus on computing the instance-optimal rule $f^*$, i.e., computing the optimal distribution in $\argmin_{p \in \Delta(\alts)} \dist(p,\prefp)$ given a preference profile $\prefp$. \citet{BCHL+15} prove that this can be accomplished in polynomial time. Specifically, they first write a \emph{nonlinear program} for the problem, which involves multiplying the probability of selecting an alternative with the utility of an agent for the alternative, both variables of the program. Then, by considering its dual program and combining it with the primal, they design a complicated \emph{linear program} (LP) whose optimal solution yields the optimal distribution and its corresponding distortion. Their full LP is provided in \Cref{app:boutilier-lp}. This LP has $O(nm^2)$ variables and $O(nm^2)$ constraints, establishing that the optimal distribution (and its corresponding distortion) can be computed in polynomial time. 

If polynomial-time computability was the sole goal, one could write the following straightforward LP given a preference profile $\prefp$.
\begin{align}
    \text{max}  \quad & \beta & \nonumber\\
    \label{inf-con1}\text{s.t.} \quad &
     \sum_{a \in \alts} p_a \cdot \sw(a,\utp) \ge \beta \cdot \max_{a \in \alts} \sw(a,\utp), & \forall \utp \in \calC(\prefp)\\
     \label{inf-con2}& \sum_{a \in \alts} p_a = 1\\
     \label{inf-con3}&p_a \ge 0, &\forall a \in \alts.
\end{align}
Variable $p_a$ denotes the probability of selecting alternative $a$ and constraint \eqref{inf-con1} requires that $\dist(p,\utp) \le 1/\beta$ under every utility profile $\utp \in \calC(\prefp)$. Thus, $1/\beta$ becomes an upper bound on the distortion of $p$ and maximizing $\beta$ yields the optimal distribution $p$ along with its distortion $1/\beta$. While this LP has \emph{uncountably many} constraints, it admits a straightforward polynomial-time separation oracle. Given values of $\beta$ and $(p_a)_{a \in \alts}$, checking whether constraints~\eqref{inf-con2} or~\eqref{inf-con3} are violated is trivial, and finding a violated constraint in \eqref{inf-con1} (if one exists) amounts to solving the following $m$ linear programs, one for each $a^* \in \alts$:
\begin{align*}
    \text{min}  \quad & \sum_{a \in \alts} p_a \cdot \left(\sum_{i \in \voters} u_{i,a}\right) - \beta \cdot \left(\sum_{i \in \voters} u_{i,a^*}\right)\hidewidth& \\
    \text{s.t.} \quad &
     u_{i,a} \ge u_{i,b}, &\forall i \in \voters, a,b \in \alts : a \succ_i b\\
     &\sum_{a \in \alts} u_{i,a} = 1, &\forall i \in \voters\\
     &u_{i,a} \ge 0, &\forall i \in \voters, a \in \alts.
\end{align*}
If the optimal value for any of these LPs is less than $0$, the corresponding utility profile identifies a violated constraint in the original LP. Thus, the original LP can be solved in polynomial time using the ellipsoid method. 

Therefore, an important contribution of \citet{BCHL+15} is that they provide a single LP
with $O(nm^2)$ variables and $O(nm^2)$ constraints to be solved, which is much faster than solving the above LP with uncountably many constraints using a separation oracle that solves $m$ LPs each with $O(nm)$ variables and constraints in each iteration. 

Our main contribution in this section is to devise a much simpler LP with only $O(nm)$ variables and $O(nm)$ constraints, and without using sophisticated techniques such as LP duality. The bedrock of our improvement is a novel structural characterization of the worst-case utility profile for a distribution $p$ on a preference profile $\prefp$, which may be of independent interest. We also use this characterization in \Cref{sec:exact-distortion}.

\subsection{Worst-case utility profiles}
\label{sec:general-worst-case}

We prove that the distortion of any distribution $p$ on any preference profile $\prefp$ is attained at some utility profile $\utp \in \calC(\prefp)$ with the following dichotomous structure.

\begin{definition}[Dichotomous Utilities]
The (unit-sum) \emph{dichotomous utility function} with respect to ranking $\pref$ over $\alts$ and $r \in [m]$ is 
\[
{\dich}_{\pref, r}(\cd) = 
{\begin{cases}
    1/r & \text{if} \; {\pref}^{-1}(\cd) \le r,
    \\
    0 & \text{o.w.};
\end{cases}
}
\]
that is, the agent is indifferent among her top $r$ alternatives and has zero utility for the remaining alternatives. We say that a utility profile $\utp$ is dichotomous if the utility function $u_i$ of each agent $i \in \voters$ is dichotomous with respect to $\pref_i$ and some $r_i \in [m]$. 
\end{definition}

We are ready to prove our structural insight.
\begin{theorem}
\label{thm:dichotomous}
For any distribution $p \in \Delta(\alts)$ and preference profile $\prefp$, there exists a dichotomous utility profile $\utp^* \in \calC(\prefp)$ where $p$ attains its worst distortion (i.e., $\dist(p,\utp^*) = \dist(p,\prefp)$).
\end{theorem}

\begin{proof}
Let $d = \dist(p,\prefp)$. If \(d=\infty\), there exists a utility profile consistent with \(\prefp\) such that some alternative has positive social welfare while \(\sw(p,\utp)=0\). Since utilities are nonnegative, for each agent with positive utility for the numerator alternative we may choose an appropriate dichotomous extreme point preserving zero expected utility under \(p\), yielding a dichotomous profile with infinite distortion. Hence assume henceforth that \(d<\infty\). Then,
\begin{align}
\max_{\utp \in \calC(\prefp)} & \; \frac{\max_{\cd \in \alts} \sw(\cd, \utp)}{\sw(p, \utp)} = d & \iff
\nonumber
\\
\max_{\utp \in \calC(\prefp)} & \; \left(\max_{\cd \in \alts} \, \sw(\cd, \utp) - d \cdot \sw(p, \utp)  \right)= 0 & \iff
\nonumber
\\
\max_{\cd \in \alts} & \; \smash{\max_{\utp \in \calC(\prefp)}} \, \sum\nolimits_{i \in \voters} \;
\left(u_i(\cd) - d \cdot u_i(p)\right) = 0 & \iff
\nonumber
\\
\max_{\cd \in \alts} & \; \sum\nolimits_{i \in \voters} \; \smash{\max_{u_i \in \calC(\pref_i)}} \; \left(u_i(\cd) - d \cdot u_i(p)\right) = 0,
\label{eq:linearization}
\end{align}
where in the last transition, the maximum can be taken over each agent separately due to the expression being linear over the agents.

It remains to show that, for any fixed alternative $\cd$ and agent $i$, $u_i(\cd) - d \cdot u_i(p)$ is maximized at some dichotomous utility function $u_i^* \in \calC(\pref_i)$. Note that the dichotomous utility functions with respect to $\pref_i$ are the extreme points of the polytope $\calC(\pref_i)$ of unit-sum utility functions consistent with $\pref_i$; equivalently, every $u_i \in \calC(\pref_i)$ can be expressed as a convex combination of these dichotomous utility functions. Hence,
\begin{align*}
&\max_{u_i \in \calC(\pref_i)} \; \left(u_i(\cd) - d \cdot u_i(p)\right)\\ 
&= \max_{\vec{\alpha} \in \Delta^m} \; \Big(\sum\nolimits_{r \in [m]} \alpha_r \cdot \dich_{\pref_i,r}(\cd) - d \cdot \sum\nolimits_{r \in [m]} \alpha_r \cdot \dich_{\pref_i,r}(p)\Big)\\
&= \max_{\vec{\alpha} \in \Delta^m} \; \sum\nolimits_{r \in [m]} \alpha_r \cdot \left(\dich_{\pref_i,r}(\cd) - d \cdot \dich_{\pref_i,r}(p)\right),
\end{align*}
where $\Delta^m = \{\vec{\alpha} \in [0,1]^m : \sum_{r \in [m]} \alpha_r = 1\}$ is the $(m-1)$-simplex. Thanks to the final expression being linear in $\vec{\alpha}$, the maximum is attained at $\alpha^*$ where $\alpha^*_{r} = 1$ (i.e., $u_i^* = \dich_{\pref_i,r}$) for some $r \in [m]$.
%
\end{proof}

Given \Cref{thm:dichotomous}, we can replace the maximum over all $u_i \in \calC(\pref_i)$ in \Cref{eq:linearization} with a maximum over dichotomous utility functions with respect to $\pref_i$ to obtain:
\begin{align}
&\dist(p, \prefp) = d \qquad \iff
\nonumber
\\
&\max_{\cd \in \alts} \; \sum_{i \in \voters} \; \max_{r \in [m]} \; \frac{1}{r} \bigg(\ind\left[\rank_i(\cd) \le r\right] - d \cdot \sum_{\ell = 1}^r p_{\pref_i(\ell)} \bigg) = 0.
\label{eq:simple-linearization}
\end{align}
We revisit \Cref{eq:simple-linearization} later to derive our results.

\subsection{A faster polynomial-time algorithm}




Building on the novel characterization of worst-case utility profiles from \Cref{thm:dichotomous}, we design a simpler linear program for computing the instance-optimal rule. This is the main result of this section.

\begin{theorem}
\label{thm:improved-lp}
Given a preference profile $\prefp$, there exists a linear program with $O(nm)$ variables, $O(nm)$ constraints, and $O(nm)$ size for computing the optimal distribution $p^* \in \argmin_{p \in \Delta(\alts)} \dist(p,\prefp)$ and its distortion $\dist(p^*,\prefp)$.
\end{theorem}
\begin{proof}
We derive the final linear program gradually via a number of insights and transformations. Let $\prefp$ be the given preference profile. First, note that the distortion $\dist(p,\prefp)$ of a distribution $p$ is the minimum value $d$ for which \Cref{eq:simple-linearization} holds. This yields the following optimization problem which computes the optimal distribution $p$ and its corresponding distortion $d$.
\begin{align*}
    \text{min}  \quad & d  \\
    \text{s.t.} \quad &
     \delta_{i,\cd} \geq  \frac{1}{r}\bigg(\ind[\rank_{i}(\cd) \le r] - d \cdot \sum\nolimits_{\ell = 1}^r p_{\pref_i(\ell)}\bigg) &
     \forall i \in \voters, \cd \in \cands, r \in [m]\\
    & \sum\nolimits_{i \in \voters} \delta_{i, a} \leq 0 &
    \forall a\in \alts\\
    &\sum\nolimits_{a \in \alts} p_a  = 1\\
    & p_a \geq 0&
    \forall a \in \alts.
\end{align*}
Here, $\delta_{i,a}$ upper bounds the expression from \Cref{eq:simple-linearization}, and by the second set of constraints we force that $\max_{a \in \alts}\sum_{i\in\voters}\delta_{i,a}\leq 0$. 
Since the right hand side of the first set of constraints is decreasing in terms of $d$, and the optimization is set to find the minimum feasible $d$, in the optimum we have
$\max_{a \in \alts}\sum_{i\in\voters}\delta_{i,a}=0$, which implements precisely \Cref{eq:simple-linearization}.

\paragraph{Linearization} However, this is not a linear program due to the multiplication of $d$ with the $p_a$ variables in the first constraint. To linearize it, we introduce a new variable $\hat{p}_a = d \cdot p_a$ for each $a \in \alts$. Since $\sum_{a \in \alts} \hat{p}_a = d$, the above program is equivalent to
\begin{align*}
    \text{min} \quad & \sum\nolimits_{\cd \in \cands} \hat p_\cd \\
    \text{s.t.} \quad &
    \delta_{i,\cd} \geq  \frac{1}{r}\bigg(\ind[\rank_{i}(\cd) \le r] - \sum\nolimits_{\ell = 1}^r \hat p_{\pref_i(\ell)}\bigg) \\
    &
    \hspace{2.8cm}
    \forall i \in \voters, \cd \in \cands, r \in [m]\\
    & \sum\nolimits_{i = 1}^n \; \delta_{i, \cd} \leq 0 
    \hspace{2.5cm} \forall \cd \in \cands\\
    & \hat p_\cd \geq 0
    \hspace{3.6cm}
    \forall \cd \in \cands
\end{align*}

This is now a linear program, whose optimal solution $\hat{p}$ yields both the optimal distortion $\sum_{a \in \alts} \hat{p}_a$ and the optimal distribution given by $p_a = \hat{p}_a/\sum_{b \in \alts} \hat{p}_b$ for all $a \in \alts$. While it already has $O(nm)$ variables, an improvement over $O(nm^2)$ variables of \citet{BCHL+15}, it still has $O(nm^2)$ constraints, the same as them.

\paragraph{Optimizing the number of constraints} To reduce the number of constraints to $O(nm)$, the key observation is the following reconstruction of the constraints that bound $\delta_{i, \pref_i(r)}$. For $i \in \voters$ and $r \in [m]$, let $s_{i, r} := \sum_{\ell = 1}^r \hat p_{\pref_i(\ell)}$. Then, the first constraint in the above program can be written as
\begin{equation}
\label{eq:toward-optimization}
\delta_{i, \pref_i(r)} \ge \max
    \Big\{\;
        \max_{\ell \in [r - 1]} -\frac{1}{\ell} \cdot s_{i, \ell}, 
        \max_{\ell \in [r, m]} \frac{1}{\ell} \cdot \left(1 - s_{i, \ell}\right)
    \,\Big\}.
\end{equation}
Define $\alpha_{i, r} = \max_{\ell \in [r]} -\frac{1}{\ell} \cdot s_{i, \ell}$ and $\beta_{i, r} = \max_{\ell \in [r, m]} \frac{1}{\ell} \cdot \left(1 - s_{i, \ell}\right)$. Then, we can bound $\delta_{i, \pref_i(r)}$ using only two constraints:
\[
\delta_{i, \pref_i(r)} \ge \alpha_{i, r - 1}, \quad \text{and} \quad \delta_{i, \pref_i(r)} \ge \beta_{i, r}.
\]
The $\alpha$'s and $\beta$'s can be set using $O(nm)$ constraints as in the final linear program below.

Linear program \PP consists of $O(nm)$ variables and $O(nm)$ constraints. The third constraint type involves $n$ variables and there are $m$ of such constraints.  All other constraints involve at most $3$ variables. Thus, the size (number of non-zero coefficients) of \PP is also $O(nm)$.
\end{proof}

\subsection{Distortion of a rule on a preference profile}\label{sec:distortion-of-rule-on-preference-profile}

While the LP above computes the instance-optimal rule $f^*$ (i.e., computes the optimal distribution $p^*$ on a given preference profile $\prefp$), in practice one may wish to implement a different voting rule $f$ due to, for example, its normative properties or the status quo. In such cases, one may wish to know the distortion of the distribution $f(\prefp)$ it returns on $\prefp$, which may not be the optimal distribution $p^*$. 

To the best of our knowledge, the only approach to solve this problem prior to our work was to use the linear program of \citet{BCHL+15} and fixing the selection probabilities to match $f(\prefp)$ (instead of letting them be variables). Adapting this approach to our optimized linear program \PP already provides a faster algorithm, but linear-programming-based algorithms are still slow. Using insights from \Cref{thm:dichotomous}, we develop an $O(nm\log(nm))$ time combinatorial algorithm for this problem which avoids solving linear programs. 

\begin{tcolorbox}
\centerline{\underline{Linear Program \PP}}
\begin{align*}
\label{eq:lp-optimized}
\text{min} \quad & \sum_{\cd \in \cands} \hat p_\cd \\
    \text{s.t.} \quad &
    \delta_{i,\pref_i(r)} \geq \alpha_{i, r - 1}
    & \forall i \in \voters, r \in [2, m]\\
    & \delta_{i, \pref_i(r)} \geq \beta_{i, r} & \forall i \in \voters, r \in [m]\\
    & \sum\nolimits_{i = 1}^n \delta_{i, \cd} \le 0 & \forall \cd \in \cands \\
    \intertext{Partial sums:}
    & s_{i, 1} = \hat p_{\pref_i(1)} & \forall i \in \voters \\
    & s_{i, r} = s_{i, r - 1} + \hat p_{\pref_i(r)} & \forall i \in \voters, r \in [2, m]\\
    \intertext{Top partial maximums:}
    & \alpha_{i, r} \ge \alpha_{i, r - 1} & \forall i \in \voters, r \in [2, m - 1]
    \\
    &\alpha_{i, r} \ge \frac{1}{r} \cdot (-s_{i, r}) & \forall i \in \voters, r \in [m - 1]\\
    \intertext{Bottom partial maximums:}
    & \beta_{i, r} \ge \beta_{i, r + 1} & \forall i \in \voters, r \in [m - 1]
    \\
    & \beta_{i, r} \ge \frac{1}{r} \left( 1 - s_{i, r} \right) & \forall i \in \voters, r \in [m]
    \\
    \intertext{Variable ranges:}
    & \hat p_\cd \ge 0 & \forall \cd \in \cands
    \\
    & \delta_{i, \pref_i(r)}, \beta_{i, r} \in \mathbb{R} & \forall i \in \voters, r \in [m]\\
    & \alpha_{i, r} \in \mathbb{R} & \forall i \in \voters, r \in [m-1]
\end{align*}%
\end{tcolorbox}%

\begin{algorithm}[t]
\SetAlgoNlRelativeSize{-1}
\KwIn{Preference Profile $\prefp$, distribution $p$, real number $d$}
\KwOut{$\texttt{Yes}$, $\texttt{No}$ with a witness utility profile}
\BlankLine
\lFor{$i \in \voters, \cd \in \cands$}{$\delta_{i, \cd} \gets 0$}
\For{$i \in \voters$}{
    $s_{i, 0} \gets 0$, $\alpha_{i,0} \gets -\infty$,
    $\beta_{i,m + 1} \gets -\infty$\;
    \lFor{$r = 1$ to $m$}{~$\;s_{i,r} \gets s_{i,r - 1} + p_{\pref_i(r)}$}
    \lFor{$r = 1$ to $m$}{
        ~$\alpha_{i,r} \gets \max\{\, \alpha_{i,r - 1}, \, -\frac{1}{r} \cdot d \cdot s_{i,r}\,\}$
    }
    \lFor{$r = m$ to $1$}{
        ~$\beta_{i,r} \gets \max\{\, \beta_{i,r + 1}, \, \frac{1}{r} \cdot \left(1 - d \cdot s_{i,r}\right)\,\}$}
    \lFor{$r = 1$ to $m$}{
        ~$\delta_{i,\pref_i(r)} \gets \max \left\{\alpha_{i,r - 1}, \beta_{i,r}\right\}$
    }
}
\eIf{$\max_{\cd \in \cands} \sum_{i}\delta_{i,\cd} \le 0$}
{
    \Return{\texttt{Yes}}
}
{
    $\cd^* \gets \arg\max_{\cd \in \cands} \sum_{i}\delta_{i,\cd}$\;
    \For{$i \in \voters$}{$r^*_i \gets \argmax\nolimits_{r \in [m]} \frac{1}{r} \left( \ind\left[\rank_i(\cd^*) \le r\right] - d\cdot s_{i, r}\right)$\;
    $u_i \gets \dich_{\pref_i, r^*_i}$\tcp*{Respective dichotomous utility}
    }
    \Return{\texttt{No}, \text{due to utility profile}~$\{u_i\}_{i \in \voters}$}
}
\caption{$\operatorname{CheckDistortion(\text{\small{of}}\; p, \,\text{\small{w.r.t.}}\; \prefp, \,\text{\small{is at most}}\; d})$}
\label{alg:check_distortion_is_at_most_d}
\end{algorithm}


\begin{theorem}
\label{thm:measure-distortion}
There is an algorithm (Algorithm~\ref{alg:measure_exact_distortion}) to compute the distortion $\dist(p, \prefp)$ of distribution $p$ on preference profile $\prefp$ in $O(nm \log (nm))$ time.
\end{theorem}


The first step is to design a fast subroutine for checking whether $\dist(p,\prefp) \le d$ for a given threshold $d$. 

\begin{restatable}{lemma}{CheckAtMostD}
\label{lem:check-at-most-d}
There is an algorithm (Algorithm~\ref{alg:check_distortion_is_at_most_d}) that, given a preference profile $\prefp$, distribution $p$, and a real number $d$, checks in linear time $O(nm)$ whether the distortion of $p$ with respect to $\prefp$ is at most $d$. If not, it finds a witness utility profile for which the distortion of $p$ is greater than $d$.
\end{restatable}


\begin{proof}
The key idea is to quickly compute \Cref{eq:simple-linearization} as both $p$ and $d$ are given. For each agent $i$ and alternative $a$, we compute the index $h_{i,a}(d)$ such that the dichotomous utility function corresponding to $\pref_i$ and $h_{i,a}(d)$ maximizes the inner expression in \Cref{eq:simple-linearization}: 

\begin{equation}
\label{eq:h_def}
h_{i, \cd}(d) = \argmax\limits_{r \in [m]} \Big\{\frac{1}{r} \cdot \Big(\ind\left[\rank_i(\cd) \le r\right] - d \cdot \sum\nolimits_{\ell = 1}^r p_{\pref_i(\ell)}\Big)\Big\}.
\end{equation}

This allows evaluating the left-hand side of \Cref{eq:simple-linearization}: if it is non-positive, the distortion is indeed at most $d$, and if it is positive, the utility profile where each agent $i$ has the aforementioned dichotomous utility function forms the sought witness.

\noindent We next present the formal proof in detail. We demonstrate that Algorithm~\ref{alg:check_distortion_is_at_most_d} accurately answers the query. According to Equation~\eqref{eq:simple-linearization}, $\dist(p, \prefp) \le d$ if and only if 
\[
\max_{\cd} \; \sum_{i \in \voters} \; \max_{r \in [m]} \; \frac{1}{r} \left(\ind\left[\rank_i(\cd) \le r\right] - d \cdot \sum_{\ell = 1}^r p_{\pref_i(\ell)} \right) \le 0.
\]
Our objective is to assess the left-hand side of the above equation. 
This is feasible if we have the value for each agent $i$ and alternative $\cd$ defined as
\begin{equation}
\label{eq:lemma1-delta-i-a}
\delta_{i, \cd} = \max_{r \in [m]} \; \frac{1}{r} \left(\ind\left[\rank_i(\cd) \le r\right] - d \cdot \sum_{\ell = 1}^r p_{\pref_i(\ell)} \right).
\end{equation}
Hence,
\[
\dist(p, \prefp) \le d
\iff
\max_{\cd} \sum_{i \in \voters} \delta_{i, \cd} \le 0.
\]
The objective above is separable by agents. Therefore, it suffices to prove that for each voter $i$, we can compute ${\delta_{i, \cd}}$ for all ${\cd \in \cands}$ in $O(m)$ time.  This is accomplished in lines 2-9 of Algorithm~\ref{alg:check_distortion_is_at_most_d}. We calculate these values similar to the linear program \PP. Let $s_{i, r} = \sum_{\ell = 1}^r p_{\pref_i(\ell)}$. These values can be computed in $O(m)$ time, as shown in line 4. Next, let
\[
\alpha_{i, r} = \max_{\ell \in [r]} -\frac{1}{\ell} \cdot d \cdot s_{i, \ell}.
\]
These values can also be computed in $O(m)$ time, as depicted in line 5. Each $\alpha_{i, r}$ is the maximum of the preceding one $\alpha_{i, r-1}$ and the new term $-\frac{1}{r} \cdot d \cdot s_{i, r}$. Similarly, define
\[
\beta_{i, r} = \max_{\ell \in [r, m]} \frac{1}{\ell} -\frac{1}{\ell} \cdot d \cdot s_{i, \ell},
\]
and compute the series in $O(m)$, as shown in line 6.  With these values available, Equation~\eqref{eq:lemma1-delta-i-a} can be computed in $O(1)$ for each $\cd \in \cands$ as follows:
\[
\delta_{i, \cd} = \max\{\alpha_{i, \rank_i(\cd) - 1}, \beta_{i, \rank_i(\cd)}\},
\]
as executed in line 7. Consequently, Algorithm~\ref{alg:check_distortion_is_at_most_d} accurately determines whether $\dist(p, \prefp) \le d$ or not.  If it exceeds $d$, we identify the candidate $\cd^* \in \cands$ that maximizes $\sum_{i \in \voters} \delta_{i, \cd}$ in $O(nm)$ time (line 11).  Having $\cd^*$ fixed, for each agent $i$, we can identify the dichotomous utility vector maximizing Equation~\eqref{eq:lemma1-delta-i-a} in $O(m)$ (using the auxiliary variable $s_{i, r} = \sum_{\ell = 1}^r p_{\pref_i(\ell)}$) and return the witness utility profile.
\end{proof}

\Cref{lem:check-at-most-d} immediately implies that one can perform a binary search for the desired distortion $d^* := \dist(p,\prefp) \in [1, \infty]$ and get $\varepsilon$-close in time $O(nm \log(d^*/{\varepsilon}))$. However, instead of searching for $d^*$ in a continuous range, we can focus on $O(nm)$ ``pivotal'' values of $d$ at which $h_{i,a}(d)$ defined in \Cref{eq:h_def} can change. Binary search over these values identifies an interval containing $d^*$, from which the exact distortion can be recovered in $O(nm)$ time. Computing these pivotal values requires additional ideas based on convex hulls (specifically, Graham's Scan~\cite{graham1972efficient}).
\\

\noindent First, we show the method of determining the pivotal values.

\begin{restatable}{lemma}{DiscreteDs}
\label{lem:discrete-ds}
There exists a set
$D=\{d_0=0<d_1<\cdots<d_T=\infty\}$ of $O(nm)$ values such that,
for every $\ell\in[T]$, $i\in\voters$, and $\cd\in\cands$,
the maximizing index $h_{i,\cd}(d)$ in \Cref{eq:h_def} is constant over
$d\in(d_{\ell-1},d_\ell)$.
The pivotal values in $D$ can be generated in
$O(nm)$ time.
\end{restatable}

\begin{proof}
Fix an agent $i$. We show that there are $O(m)$ pivotal values
for $i$ and that they can be generated in $O(m)$ time.
Write $\cd=\pref_i(r)$ and define
\[
s_{i,k}:=\sum_{q=1}^k p_{\pref_i(q)}.
\]
As in Equation~\eqref{eq:toward-optimization}, we split the maximum
in \Cref{eq:h_def} into the cases $k<r$ and $k\ge r$.

For the first case ($k<r$), the maximizing index is independent
of $d$ for $d>0$, since $d$ is a common positive multiplicative
factor. Indeed,
\[
\max_{k<r}
\left\{-\frac{s_{i,k}}{k}d\right\}
=
-\gamma_{i,r}d,
\qquad
\gamma_{i,r}:=
\min_{k<r}\frac{s_{i,k}}{k}.
\]
When $r=1$, this case is absent.

For the second case ($k\ge r$), define
\[
f_{i,r}(d)
:=
\max_{k\in[r,m]}
\left\{
\frac{1}{k}-\frac{s_{i,k}}{k}d
\right\}.
\]
Thus, on $d\ge0$, $f_{i,r}$ is the upper envelope of the lines
\[
l_{i,k}(d)
:=
\frac{1}{k}-\frac{s_{i,k}}{k}d,
\qquad
L_{i,r}:=\{l_{i,k}:k\in[r,m]\}.
\]

We construct these envelopes incrementally, starting with
$L_{i,m}$ and adding $l_{i,r}$ for $r=m-1,\ldots,1$.
Since
\[
l_{i,r}(0)=\frac{1}{r}>\frac{1}{k}=l_{i,k}(0)
\qquad\text{for every }k>r,
\]
each new line lies strictly above the existing envelope at $d=0$.
The existing envelope is convex, so its difference with the new
line is convex and strictly negative at $d=0$. Consequently, the
new line can intersect the existing envelope at most once for
$d>0$. Each update therefore replaces an initial portion of the
existing envelope with the new line.

We implement these updates using a stack, with the segment
closest to $d=0$ at the top. When adding a line, we pop the
segments it completely covers, shorten the first surviving
segment if necessary, and push the new segment. We discard
segments of zero length. Each line is pushed once and popped at
most once, and each insertion shortens at most one surviving
segment. Hence, the total update time is $O(m)$.
Each insertion creates at most one new transition point, which
we record even if a later insertion removes it from the envelope.
Thus, the envelopes of all sets $L_{i,r}$ together have only
$O(m)$ distinct transition points, and these points can be
generated in $O(m)$ time.

It remains to account for switches between the two cases.
Let $\mu_i:=\min_{k\in[m]}s_{i,k}/k$, and let $t_i$ be the
largest index attaining this minimum. For $1<r\le t_i$,
the line $l_{i,t_i}(d)$ lies strictly above
$-\gamma_{i,r}d$ for all $d\ge0$, since
$\mu_i\le\gamma_{i,r}$. Thus, no switch occurs.

For $r>t_i$, we have $\gamma_{i,r}=\mu_i$ and
$s_{i,k}/k>\mu_i$ for every $k\ge r$. Each line $l_{i,k}$
therefore crosses $-\mu_i d$ once and remains below it
afterwards. Hence, the unique switch occurs at
\[
x_{i,r}
=
\max_{k\ge r}\frac{1}{s_{i,k}-k\mu_i}
=
\frac{1}{\min_{k\ge r}(s_{i,k}-k\mu_i)}.
\]
All these values can be computed in $O(m)$ time using suffix
minima, and we add them to the pivotal values.

Together with the envelope transitions, these crossings give
$O(m)$ pivotal values for agent $i$, generated in $O(m)$ time.
Taking their union over all agents and adding $0$ and $\infty$
gives the required set $D$ in $O(nm)$ time.
\end{proof}

\begin{algorithm}[htb!]
\SetAlgoNlRelativeSize{-1}
\KwIn{Preference profile $\prefp$, distribution $p$}
\KwOut{Distortion value of $p$ w.r.t. $\prefp$}
\BlankLine
$D \gets [0,\infty]$\;
\For{$i \in \voters$}{
    $s_{i,0} \gets 0$\;
    \lFor{$r=1$ to $m$}{
        $s_{i,r} \gets s_{i,r-1}+p_{\pref_i(r)}$
    }
    $\mu_i \gets \min_{k\in[m]} s_{i,k}/k$\;
    \tcc{Construct the upper envelopes of
    $L_{i,r}=\left\{l_{i,k}(d)=\frac{1}{k}-d \cdot \frac{s_{i,k}}{k} \mid k\in[r,m]\right\}$
    on $d\ge0$, and record their transition points and crossings
    with the first-case line.}    $b \gets \infty$, $\operatorname{upper-hull}_i \gets \emptyset$\;
    \For{$r=m$ down to $1$}{
        $b \gets \min\{b,\ s_{i,r}-r\mu_i\}$\;
        Add the line
        $l_{i,r}(d)=\frac{1}{r}-\frac{s_{i,r}}{r}d$
        to $\operatorname{upper-hull}_i$ on $d\ge0$
        using the stack update in \Cref{lem:discrete-ds}\;
        $d'_{i,r} \gets$ the right endpoint of the new line's
        segment in $\operatorname{upper-hull}_i$
        \tcp*{Possibly $\infty$}
        \lIf{$d'_{i,r}<\infty$}{
            Append $d'_{i,r}$ to $D$
        }
        \lIf{$r>1$ and $b>0$}{
            Append $1/b$ to $D$
        }
    }
}
Sort $D$ in increasing order and remove duplicates\;
Using the checker of \Cref{lem:check-at-most-d}, perform binary
search to find consecutive values $d_{\ell-1},d_\ell\in D$
such that $\dist(p,\prefp)>d_{\ell-1}$ and
$\dist(p,\prefp)\le d_\ell$\;
Choose any $d\in(d_{\ell-1},d_\ell)$ and, for every
$\cd\in\cands$, find the corresponding maximizing dichotomous
utility profile $\utp^\cd$
\;
$d^* \gets \max_{\cd\in\cands}
\frac{\sw(\cd,\utp^\cd)}{\sw(p,\utp^\cd)}$\;
\Return{$d^*$}\;
\caption{$\operatorname{MeasureDistortion}
(\text{\small of}\;p,\,
\text{\small w.r.t.}\;\prefp)$}
\label{alg:measure_exact_distortion}
\end{algorithm}

Using \Cref{lem:check-at-most-d,lem:discrete-ds}, we can prove \Cref{thm:measure-distortion}.
\begin{proof}[Proof of \Cref{thm:measure-distortion}]
By \Cref{lem:discrete-ds}, we can find a set $D$ of $O(nm)$ values such that, for any two consecutive values $d_{\ell-1}, d_{\ell} \in D$ and every alternative $\cd$, the corresponding maximizing dichotomous utility profile stays unchanged across all $d \in (d_{\ell-1}, d_{\ell})$. By performing a binary search on this set of values using \Cref{lem:check-at-most-d}, we can find two consecutive values $d_{\ell-1},d_\ell$ such that $\dist(p,\prefp)>d_{\ell-1}$ and $\dist(p,\prefp)\le d_\ell$. Then, we know that $d^* \in (d_{\ell-1},d_\ell]$.

Fix any $d \in (d_{\ell-1},d_\ell)$, and for every alternative $\cd$, let $\utp^\cd$ be its corresponding maximizing dichotomous utility profile. Since these profiles stay unchanged throughout this interval, \Cref{eq:simple-linearization} is equal to
\[
\max_{\cd\in\cands}
\left\{
\sw(\cd,\utp^\cd)-d\cdot\sw(p,\utp^\cd)
\right\}.
\]
Therefore, its zero threshold, which is precisely the exact distortion, is
\[
d^*
=
\max_{\cd\in\cands}
\frac{\sw(\cd,\utp^\cd)}
     {\sw(p,\utp^\cd)}.
\]
This also covers the case $d^*=d_\ell$ by continuity. The $O(nm\log(nm))$ running time follows from constructing and sorting the $O(nm)$ pivotal values, $O(\log(nm))$ calls to the $O(nm)$ time algorithm from \Cref{lem:check-at-most-d} as part of the binary search, and an additional $O(nm)$ time to compute the final distortion.
\end{proof}

\section{Distortion of randomized positional scoring rules}\label{sec:exact-distortion}
In the previous section, we were concerned with computing the distortion of a given distribution or the optimal distribution on a \emph{given} preference profile. However, the (overall) distortion of a rule requires computing its worst-case distortion across all preference profiles --- a significantly more complex task, for which prior work provides no algorithms to the best of our knowledge. Our main contribution in this section is to devise such an algorithm for the following well-studied family of rules, and we do so by solving a non-convex quadratic program. Note that such programs, including ours, are not known to be solvable in polynomial time.

\subsubsection*{Randomized positional scoring rules.} Given a scoring vector $\sv = (s_1, s_2, \ldots, s_m)$, the \emph{positional scoring rule} (PSR) assigns a score of $s_j$ to an alternative each time it appears at the $j$-th position in the ranking of any agent (for each $j$), and (deterministically) selects an alternative with the highest total score. A \emph{randomized positional scoring rule} (RPSR) $f_{\sv}$ assigns scores to alternatives in the same way, but selects each alternative with probability proportional to its score. On preference profile $\prefp$, this selects each alternative $a \in \alts$ with probability
$\prs(a, \prefp) := \Pr[f_{\sv}(\prefp) = a] =  \frac{1}{n|\sv|} \sum_{i \in \voters} s_{\rk_i(a)}$.

These rules, also called point-voting schemes~\cite{barbera1978nice}, are known for their strategyproofness~\cite{gibbard1977manipulation}. \citet{ebadian2023explainable} provide asymptotic distortion bounds for many rules in this family. 

In this section, we first devise an algorithm for computing the (exact) distortion of a given RPSR. Then, we iteratively use our algorithm, together with an appropriately constructed linear program, to find the RPSR with the smallest distortion for a given number of alternatives $m$. Finally, we present the bounds achieved via our techniques for up to $m=5$ alternatives, for both the best RPSR and commonly used RPSRs from the literature. 

\subsection{Worst-case utility and preference profiles}\label{sec:qp-worst-case-instances}

Let $\alts = \{a_1,a_2,\ldots,a_m\}$. Let us define three properties of a pair consisting of a preference profile $\prefp$ and a utility profile $\utp \in \calC(\prefp)$. 

\begin{properties}[label=(P\arabic*)]
    \item \label{r-dichotom} $\utp$ is dichotomous.
    \item \label{sw-order} $\sw(a_j,\utp) \geq \sw(a_{j+1}, \utp)$ for all $j \in [m-1]$.
    \item \label{eq-order} For each agent $i \in \voters$ and alternatives $a_j,a_{j'} \in \alts$ with $j < j'$, if $u_i(a_j) = u_i(a_{j'})$, then $a_{j'} \succ_i a_j$.
\end{properties}

\Cref{r-dichotom} is our insight from \Cref{thm:dichotomous} that the worst-case utility profile $\utp$ is dichotomous WLOG. \Cref{sw-order} uses neutrality of RPSRs (i.e., that they do not depend on the names of the alternatives) to further restrict the search space for $\utp$. Unlike in \Cref{sec:optimal-distribution} where the preference profile $\prefp$ was given, here we seek to identify its worst case, for which \Cref{eq-order} is our novel insight. 

Let $W_m = \set{(\pref_1,u_1),\ldots,(\pref_t,u_t)}$ denote the set of possible pairs of preference ranking $\pref_i$ and utility function $u_i \in \calC(\pref_i)$ that any agent can have in any pair of preference profile $\prefp$ and utility profile $\utp \in \calC(\prefp)$ satisfying \Cref{r-dichotom,sw-order,eq-order}; note that $t := |W_m|$ and from here on, with slight abuse of notation, we will use $i$ to index a pair (agent type) in $W_m$ instead of denoting an individual agent. The next lemma shows that the three properties above significantly reduce the search space.

\begin{restatable}{lemma}{lemSizeOfSet}
    $|W_m| \le 2^m-2$.
    \label{lem:final}
\end{restatable}

\begin{proof}
    Since utility profiles are dichotomous, each $u_i$ partitions the alternatives into two sets with equal utility. We know that, within each of these two sets, alternatives appear in decreasing order of their indices. That means if we know the members of each set then $\pref_i$ is unique. So the number of different pairs of $(\pref_i, u_i)$ is equal to the number of ways in which we can partition the alternatives into two sets. In addition we have to make sure that the first set (alternatives with $1/r$ utility) is nonempty. That means we have at most $2^m-1$ such pairs. In addition we know that if $a^*$ is the only member of the bottom set (the only one with zero utility), then by changing the utility function from $\dich_{m-1}$ to $\dich_m$ we get a higher distortion. That means the set of all possible $(\pref_i, u_i)$ includes at most $2^m-2$ members.
\end{proof}

\Cref{tab:three} 
shows the $2^3-2 = 6$ possible pairs of preference ranking and utility function that we need to consider for $m=3$ alternatives. Finally, we show that every RPSR achieves its worst distortion on a pair $(\prefp,\utp)$ satisfying \Cref{r-dichotom,sw-order,eq-order}. 

\begin{restatable}{lemma}{lemHaveProperties}
    For every randomized positional scoring rule $f_{\sv}$, there exists a pair $(\prefp,\utp)$ of preference and utility profiles satisfying \Cref{r-dichotom,sw-order,eq-order} at which $f_{\sv}$ attains its worst distortion.
    \label{lem:HaveProperties}
\end{restatable}

\begin{proof}
    First by \Cref{thm:dichotomous} we know that a worst-case instance with dichotomous utility profile exists. We can compute the social welfare of the alternatives in that instance and relabel the alternatives to match the ordering $\succ$. That means we have a worst case instance that satisfies \Cref{r-dichotom,sw-order}. 
    We want to prove that there exists an instance that satisfies \Cref{eq-order}. For the sake of contradiction assume that such instance does not exist. Consider all worst-case instances that satisfy \Cref{r-dichotom,sw-order} but not \Cref{eq-order}, and among all those instances let $(\utp, \prefp)$ be the one with the minimum number of agents $i \in \voters$ and pairs $a, b$ with $\sw(a, \utp) \geq \sw(b, \utp)$ such that $u_i(a) = u_i(b)$ but $a \succ_i b$. If such pair exists in an agent's ranking then a consecutive pair of alternatives in the ranking with such condition also exists. So w.l.o.g. assume that $\rank_i(a) = \rank_i(b) - 1$.
    Let this instance be $(\utp, \prefp)$. We want to create instance $(\utp', \prefp')$ with (at least) the same distortion with less number of such pairs.
    
    Define $r_a := \rank_i(a)$ and $r_b := \rank_i(b)$.
    If we switch the position of $a$ and $b$ in $\pref_i$
    to form $\prefp'_i$, we do not change the social welfares and we have
    \begin{align*}
        &\dist(f_{\sv}(\prefp'), \utp') = \frac{\sw(a^*, \utp')}{\sum\limits_{a' \in \alts} \prs(a', \prefp') \sw(a', \utp')}\\
        &= \frac{\sw(a^*, \utp)}{\sum\limits_{a' \in \alts} \prs(a', \prefp') \sw(a', \utp) + \frac{1}{|\vec s|}(s_{r_b}-s_{r_a})\sw(a, \utp) + \frac{1}{|\vec s|}(s_{r_a}-s_{r_b})\sw(b, \utp)}\\
        &= \frac{\sw(a^*, \utp)}{\sum\limits_{a' \in \alts} \prs(a', \prefp') \sw(a', \utp) + \frac{1}{|\vec s|}(s_{r_a}-s_{r_b})(\sw(b, \utp)-\sw(a, \utp))}
        \\&\geq \dist(f_{\sv}(\prefp), \utp).
    \end{align*}
    This has one less ``bad'' pair since we do not change the ranking of other voters and the ranking of other alternatives in voter $i$'s ranking.
\end{proof}

\subsection{Computing the distortion of an RPSR}\label{sec:qp:worst-case-distortion-given-rpsr}
Fix any RPSR $f_{\sv}$. For $i \in [t]$, define $q_i$ to be the fraction of agents who have preference ranking and utility function $(\pref_i, u_i) \in W_m$. Note that the $1\times t$ vector $q = (q_1,\ldots,q_t)$ satisfies $\sum_{i \in [t]} q_i = 1$ and can capture any worst-case instance $(\prefp,\utp)$ satisfying \Cref{r-dichotom,sw-order,eq-order}, for any number of agents $n$. 

We want to understand the distortion of $f_{\sv}$ at a given $q$ and then write a program to optimize over $q$. Let $U_{m \times t}$ be a matrix where $U_{j,i}=u_i(\cd_j)$, and $P_{m \times t}$ be a matrix where $P_{j, i}$ is the score that candidate $\cd_j$ gets from any agent with preference ranking $\pref_i$, i.e. $P_{j, i} = s_{\rank_i(\cd_j)}$. Also, define $C_{t \times t} = \frac{1}{|\sv|} U^\tran P$. For example, when $m=3$, matrices $U$ and $P$ corresponding to the set $W_m$ from \Cref{tab:three} are shown below. 

\begin{table}[htb!]
\small
\centering
\begin{tabular}{@{}l@{\hspace{0.25em}}|c@{\hspace{0.5em}}c@{\hspace{0.5em}}c@{\hspace{0.5em}}c@{\hspace{0.5em}}c@{\hspace{0.5em}}c@{\hspace{0.5em}}c@{}}\toprule
&$(\pref_1,u_1)$ &$(\pref_2,u_2)$ &$(\pref_3,u_3)$ &$(\pref_4,u_4)$ &$(\pref_5,u_5)$ &$(\pref_6,u_6)$\\\midrule
$\pref_i$ &$a_1$$\succ$$a_2$$\succ$$a_3$ &$a_2$$\succ$$a_1$$\succ$$a_3$ &$a_2$$\succ$$a_3$$\succ$$a_1$ &$a_3$$\succ$$a_1$$\succ$$a_2$ &$a_3$$\succ$$a_2$$\succ$$a_1$ &$a_2$$\succ$$a_3$$\succ$$a_1$ \\[0.1cm]
$u_i$ &$\left[1,\,\,  0,\,\,  0\right]$ &$\left[\frac{1}{2},\,\,  \frac{1}{2},\,\,  0\right]$ &$\left[1,\,\,  0,\,\,  0\right]$ &$\left[\frac{1}{2},\,\,  \frac{1}{2},\,\,  0\right]$ &$\left[1,\,\,  0,\,\,  0\right]$ &$\left[\frac{1}{3},\,\,  \frac{1}{3},\,\,  \frac{1}{3}\right]$ \\
\bottomrule
\end{tabular}
\caption{The set $W_3 = \set{(\pref_1,u_1),\ldots,(\pref_6,u_6)}$ from \Cref{lem:final}.}
\label{tab:three}
\end{table}

\begin{equation*}
    U \!=\! \begin{bmatrix}
1 & \frac{1}{2} & 0 & \frac{1}{2} & 0 & \frac{1}{3}\\
0 & \frac{1}{2} & 1 & 0           & 0 & \frac{1}{3}\\
0 & 0           & 0 & \frac{1}{2} & 1 & \frac{1}{3}
\end{bmatrix}\  \text{and} \   P\!=\! \begin{bmatrix}
s_1 & s_2 & s_3 & s_2 & s_3 & s_3\\
s_2 & s_1 & s_1 & s_3 & s_2 & s_1\\
s_3 & s_3 & s_2 & s_1 & s_1 & s_2
\end{bmatrix}.
\end{equation*}
\bigskip 

From the above, we can see that if $q$ represents the instance $(\prefp,\utp)$, then $\sw(\cd_j,\utp) = n \cdot q \cdot U_j^\tran$ and $\prs(\cd_j,\prefp) = \frac{1}{|\vec s|}\cdot q \cdot P_j^\tran$, where $U_j$ and $P_j$ denote the $j$-th row of $U$ and $P$, respectively. This means 
\begin{align*}
\dist(f_{\sv}) &= \frac{q \cdot U_1^\tran}{\sum_{j \in [m]} (q \cdot U_j^\tran) (\frac{1}{|\vec s|}\cdot q \cdot P_j^\tran)}
= \frac{q \cdot U_1^\tran}{ q\left(\frac{1}{|\vec s|}\sum_{j \in [m]} U_j^\tran \cdot P_j\right) q^\tran} = \frac{ q U_1^\tran}{q C q^\tran}.
\end{align*}

Hence, we arrive at the following optimization program to compute the distortion of the given RPSR $f_{\vs}$.
\begin{align*}
    \text{max} \quad & \frac{q U_1^\tran}{q C q^\tran} & \\
    \text{s.t.} \quad 
    & q \cdot U_j^\tran \leq q \cdot U_1^\tran  & \forall j \in [m] \\
    & {\textstyle\sum_{i \in [t]}}\, q_i  = 1 \\
    & q_i \geq 0 & \forall i \in [t].
\end{align*}

To transform this optimization program to a form more amenable to solving via standard solvers, we add a variable $D$ to capture the inverse of the distortion, i.e., $1/\dist(f_{\sv})$. The new objective is to minimize $D$, and we add a quadratic constraint $D \cdot (q U_1^\tran) \ge q C q^\tran$ to ensure that $D$ remains an upper bound on the inverse of the distortion. The resulting optimization program has a linear objective, a set of linear constraints, and a single quadratic constraint. 

The inverse of the optimal objective value of Program \QQ yields the distortion $\dist(f_{\sv})$ of 
a given RPSR $f_{\sv}$ as well as the instance $(\prefp,\utp)$ where the worst distortion is attained; the latter part will be useful in our algorithm in \Cref{sec:optimal_vec} for finding the best RPSR.

We remark that Program \QQ is not convex, and thus not known to be polynomial-time solvable, even to a given precision. Still, for relatively small values of $m$ (e.g., up to $m=5$), standard solvers are able to provide globally optimal solutions within reasonable running times as we show in \Cref{sec:small_vec}. 

\begin{tcolorbox}
\centerline{\underline{Quadratic Program \QQ}}\vspace{-.4cm}
\begin{align*}
    \text{min.} \quad & D & \\
    \text{s.t.} \quad 
    & D \sum\limits_{i \in [t]} q_i \cdot U_{1, i} \geq \sum\limits_{i \in [t]} \sum\limits_{k \in [t]} q_i\cdot q_k \cdot C_{i, k} & \\
    & \sum\limits_{i \in [t]} q_i \cdot U_{j, i} \leq \sum\limits_{i \in [t]} q_i \cdot U_{1, i} & \forall j \in [m] & \\
    & {\textstyle\sum_{i \in [t]}}\, q_i = 1 \\
    & q_i \geq 0  & \forall i \in [t]. &
\end{align*}
\end{tcolorbox}
\subsection{Finding the best possible RPSR}\label{sec:optimal_vec}
In this section, we make use of our quadratic program \QQ to find the RPSR that achieves the minimum distortion (and that distortion itself) for a given number of alternatives $m$. The idea is to use program \QQ repeatedly to search over the space of all RPSRs. Since this is a vast search space and solving the non-convex program \QQ takes time, we employ the technique of \citet{filos2014truthful}, which uses zero-sum games (and as a result, linear programs) to aid the search and quickly converge to the optimal solution without exploring too many points in the search space. 

\subsubsection*{Lower bounds via zero-sum games} \citet[Theorem 2]{filos2014truthful} show that any  RPSR for $m$ alternatives is a convex combination of rules $F_m^k$ for $k \in [m]$: rule $F_m^k$ (a) selects an agent uniformly at random and (b) selects one of her $k$ most-preferred alternatives uniformly at random. This result of \cite{filos2014truthful} is in fact almost a direct corollary of a result of \citet{barbera1978nice}, which was in turn obtained from the characterization of \citet{gibbard1977manipulation} of truthful rules. 

By applying Yao's minimax principle \cite{yao1977probabilistic}, we can effectively transform the design of the optimal RPSR into a (series of) zero-sum game(s) $\mathcal{G}$. In a game $\mathcal{G}$, the pure strategies of the maximizer are the rules $F_m^k$ for $k \in [m]$ and the pure strategies of the minimizer are a set of instances $\mathcal{I}$ (the construction of this set is explained in the next paragraph), where an instance is a pair of preference and utility profiles $(\prefp,\utp)$ such that $\utp \in \calC(\prefp)$. When the maximizer chooses rule $F_m^j$ and the minimizer chooses instance $(\prefp,\utp) \in \mathcal{I}$, the reward in the corresponding cell of the game is the inverse of the corresponding distortion, i.e., $1/\dist(F_m^j(\prefp),\utp)$. From the characterization of \citet{filos2014truthful}, the set of mixed strategies of the maximizer is precisely the set of RPSRs. By computing an optimal mixed strategy for the maximizer, we obtain an RPSR with the smallest distortion in the worst case over all instances in $\mathcal{I}$ (and this distortion value is the inverse of the value of game $\mathcal{G}$). Game $\mathcal{G}$ can be solved by a standard formulation of zero-sum games as linear programs. 

\vspace{0.5em}
\noindent\emph
{An iterative algorithm.} Throughout our algorithm, we maintain (a) the current distortion upper bound $d^{u}$, (b) the current distortion lower bound $d^{\ell}$, (c) a set of ``bad'' instances $\mathcal{I}$, and (d) a candidate RPSR $f_c$. Initially $d^{u}=\infty$, $d^{\ell}=0$,  $\mathcal{I}=\emptyset$, and $f_c$ is any RPSR.  

In each iteration, we run the quadratic program \QQ on the rule $f_c$, which returns both $\dist(f_c)$ and the worst-case instance $(\prefp,\utp)$ at which this distortion is attained. We update $d^{u} \gets \min(d^{u},\dist(f_c))$ and add $(\prefp,\utp)$ to $\mathcal{I}$. Then, we construct the matrix game $\mathcal{G}$ described above with the pure strategies of the maximizer being the rules $F_m^k$ and the pure strategies of the minimizer being the instances in $\mathcal{I}$. We update $d^{\ell}$ to be the inverse of the value of this game $\mathcal{G}$ and $f_c$ to be the optimal mixed strategy of the maximizer in $\mathcal{G}$. We repeat this process until a stopping criterion has been met, which is that $d^{u}-d^{\ell} \leq \varepsilon$ for a sufficiently small $\varepsilon$; the choice of $\varepsilon$ in our experiments is described in the next section. During this process, we keep track of the RPSR that achieves the distortion of $d^u$, and at termination, output that as our estimate of the best RPSR. Note that $d^u$ is the exact distortion of this rule, and thus, a valid upper bound on the distortion of the best RPSR. Similarly, $d^{\ell}$ is also a valid lower bound on the distortion of any (and thus the best) RPSR. 

\subsection{Bounds for a small number of alternatives}
\label{sec:small_vec}
By deploying the techniques presented in the previous two subsections, we are able to compute the exact distortion of several well-known RPSRs and of the best RPSRs for $m \in \set{2,3,4,5}$ alternatives.
In more detail, we employ the quadratic program \QQ that we develop in \Cref{sec:qp:worst-case-distortion-given-rpsr} to compute the exact distortion of the following well-known RPSRs:
\begin{itemize}[leftmargin=*]
\item[-] \emph{Randomized Plurality}, with scoring vector $\sv = (1, 0, \ldots 0)$,
\item[-] \emph{Randomized Borda}, with scoring vector $\sv=(m-1, \ldots, 1,0)$,
\item[-] \emph{Randomized $k$-Approval}, with scoring vector $\sv=(1,\ldots,1, 0, \ldots, 0)$, with $k$ ones, for $k=2$ and $k=3$,
\item[-] \emph{Randomized Veto}, with scoring vector $\sv = (1, 1, \ldots 1, 0)$,
\item[-] \emph{Randomized Harmonic}, with scoring vector $\sv = (1, 1/2, \ldots, 1/m)$,
\item[-] the \emph{``Golden Rule''}~\citep{BCHL+15} with scoring vector $\sv = (1, 1/2, \ldots, 1/m) + (1/m, \ldots, 1/m)$. The Golden Rule has the smallest \emph{asymptotic} distortion among all RPSRs~\cite{bhaskar2018truthful}. 
\end{itemize}
Using the iterative algorithm described in \Cref{sec:optimal_vec}, with stopping criterion $d^{u}-d^{\ell} \leq 0.005$,  we also obtain (essentially) tight bounds on the distortion of the best RPSR for $m$ alternatives. To solve the quadratic program \QQ, we use the Gurobi Optimization Solver. We remark that the main computational bottleneck of this technique is solving the quadratic program; the aided search of \Cref{sec:optimal_vec} terminated in at most 9 iterations in all cases. All computations were performed using an Apple M2 CPU with 24 GB of RAM.

\subsection{Computing the distortion of deterministic PSRs}\label{app:deterministic-PSR}
We complement these results with bounds on the best (deterministic) PSRs to demonstrate the advantages of randomization. To calculate these bounds, we first show that the (exact) distortion of a given PSR can be computed by a linear program. Here, we cannot use the zero-sum-game-aided search from \Cref{sec:qp:worst-case-distortion-given-rpsr}, so we perform a simple grid search on the space of PSRs to find the best PSR; this is still fast enough because we are now solving linear (rather than quadratic) programs to compute the distortion of PSRs.

To find the distortion of a given PSR, we take the same route as in \Cref{sec:exact-distortion} by identifying some properties for the worst-case instances. 

\begin{lemma}\label{lem:DPSR-properties}
For any (deterministic) positional scoring rule $f_{\sv}$ and alternatives $a^*, a^+$, there exists a pair consisting of a preference profile $\prefp$ and a utility profile $\utp \in \calC(\prefp)$ that satisfies the following properties and has the maximum possible distortion for that rule. 
\begin{properties}[label=(\roman*)]
    \item $a^*$ is the optimal alternative and $a^+$ is the winner.
    \item \label{det-1} For $i \in \voters$ if $a^* \succ_i a^+$ then $u_i = \dich_{\pref_i,\rank_i(a^*)}$, and $\rank_i(a^+) = \rank_i(a^*) + 1$ .
    \item \label{det-2} For $i \in \voters$ if $a^+ \succ_i a^*$ then $u_i = \dich_{\pref_i,m}$, and $\rank_i(a^+) = 1$.
\end{properties}
In addition, the number of possible pairs of $\pref, u$ that appear in one such preference and utility profile is at most $2(m-1)!$.
\end{lemma}
\begin{proof}
    For any worst-case profile, the winner and the optimal are different alternatives. Relabel the alternatives to make $a^*$ the optimal and $a^+$ the winner. Let $(\pref, \utp)$ be such a worst-case instance. We want to build $(\pref', \utp')$ with the desired properties.
    
    Consider any $i \in \voters$ where $a^* \succ_i a^+$. The distortion is \[\frac{\sw(a^*, \utp)}{\sw(a^+, \utp)},\] so to maximize the distortion we can set $u'_i = \dich_{\pref_i,\rank_i(a^*)}$. After this, if we move $a^+$ up to be just below $a^*$ we do not increase its social welfare and do not decrease its score. That means the winner remains the same and the distortion does not change, and hence we have a worst-case instance with \Cref{det-1}.

     Consider any $i \in \voters$ where $a^+ \succ_i a^*$. The distortion is \[\frac{\sw(a^*, \utp)}{\sw(a^+, \utp)},\] so to maximize the distortion we can set $u'_i = \dich_{\pref_i,m}$. After this, if we move $a^+$ to the top of the list, we do not increase its social welfare and do not decrease its score. That means the winner remains the same and the distortion does not change, and hence we have a worst-case instance with \Cref{det-2}.

    Now consider any agent in $(\pref', \utp')$. There are $(m-1)!$ rankings over $\alts \setminus \set{a^+}$. For each of them, either $a^* \succ_i a^+$, in which case by \Cref{det-1}, $u_i = \dich_{\pref_i,\rank_i(a^*)}$, and $\rank_i(a^+) = \rank_i(a^*) + 1$, or 
     $a^+ \succ_i a^*$, in which case by \Cref{det-2}, $u_i = \dich_{\pref_i,m}$, and $\rank_i(a^+) = 1$. That means we have at most two possible total-ordering-and-utility-function pairs for each of the $(m-1)$ partial orderings over $\alts \setminus \set{a^+}$.
\end{proof}

Now let $W_m = \set{(\pref_i, u_i)}$ be the set of all possible preferences and utility functions in a worst-case instance that satisfy the properties of \Cref{lem:DPSR-properties},  and define $q$ to be a vector of size $w := |W_m|$, where $q_i$ is proportional to the fraction of agents with preference ranking $\pref_i$, and utility function $u_i$. We can use the following linear program to find the distortion of this rule.

\begin{tcolorbox}
\centerline{\underline{Linear Program \PP'}}
\begin{align*}
\label{det-lp}
\text{max} \quad & \sum_{i \in [w]} q_i u_i(a^*) \\
    \text{s.t.} \quad &
    \sum_{i \in [w]} q_i s_{\rank_i(a)} \leq \sum_{i \in [w]} q_i s_{\rank_i(a^+)}
    & \forall a \in \alts\\
    &\sum_{i \in [w]} q_i u_i(a) \leq \sum_{i \in [w]} q_i u_i(a^*)
    & \forall a \in \alts\\
    &\sum_{i \in [w]} q_i u_i(a^+) = 1\\
    & q_i \geq 0 & \forall i \in [w]
\end{align*}
\end{tcolorbox}

To find the best possible scoring vector we perform a grid search on the space of possible PSRs; while this search space is very large (in fact continuous), the fact that we only have to solve a linear program for each candidate point allows us to use a rather fine grid for our search. This technically only gives us upper bounds, but if there is a better rule, it would have to lie at a point that is not on our grid. Of course, we have not proved that the distortion function is continuous, although all of our empirical observations seem to suggest that. In that case, we would expect our upper bounds to be within $\varepsilon$ of the distortion of the best possible PSR, where $\varepsilon$ would depend on the size of the grid; we take the grid granularity to be $0.001$, so the differences in the distortion between the lower and upper bounds would be insignificant. It would be interesting to show the continuity of the distortion objective, but since it is only relevant for this complementary part of our work, we leave that as a future direction. 

We remark that the best possible bounds are achieved by multiple scoring vectors. In fact, for $m=2$, \emph{any} scoring vector achieves the best distortion of $3$. The proven distortion and one of the scoring vectors that achieve that distortion for up to $m=5$ are shown in \Cref{tab:optimal_rules,tab:optimal_rules_2}.

\begin{table}[ht!]
\renewcommand{\arraystretch}{1.5}
    \centering
    \begin{tabular}{|c|c|c|} \hline 
         $m$&  Optimal PSR&  Optimal PSR Distortion  \\ \hline 
         2& $[1, 0]$ &  $3$  \\ \hline 
         3&  $[1, 0, 0]$&  $5.5$  \\ \hline 
         4& $[0.65, 0.26, 0.09, 0.0]$ &  $8.333$    \\ \hline 
         5& $[0.634, 0.22, 0.082, 0.036, 0.028]$ 	  &  $11.418$ \\ \hline
    \end{tabular}
    \caption{Optimal distortion of PSRs and their corresponding scoring vectors for $m$ up to $5$.}
    \label{tab:optimal_rules}
\end{table}

\begin{table}[ht!]
\renewcommand{\arraystretch}{1.5}
    \centering
    \begin{tabular}{|c|c|c|c|} \hline 
         $m$&  Optimal RPSR& Optimal RPSR  Distortion &\# Iterations\\ \hline 
         2 & $[0.843, 0.157]$ &  $1.5$  &$4$\\ \hline 
         3 & $[0.776, 0.189, 0.035]$ &   $1.870$&$6$\\ \hline 
         4& $[0.7247, 0.2025, 0.069, 0.0038]$ & $2.188$    &$7$\\ \hline 
         5&  $[0.6645, 0.2121, 0.0881, 0.0353, 0.0]$ &   $2.474$ & $9$\\ \hline
    \end{tabular}
    \caption{Optimal distortion of RPSRs and their corresponding scoring vectors for $m$ up to $5$.}
    \label{tab:optimal_rules_2}
\end{table}

\subsection{Lower bounds for randomized rules}
To assess how close the best RPSR is to the best rule overall, we also present lower bounds on the distortion of \emph{any} randomized voting rule. To compute these, we employ a gradient-descent style search over the space of preference profiles, employing our algorithm in \Cref{sec:distortion-of-rule-on-preference-profile} to find the best distortion on each preference profile (see Algorithm~\ref{alg:measure_exact_distortion}).

As defined in \Cref{sec:small_vec}, each worst-case instance is identified by $q$, a unit sum vector of size $t$. Here $t$ is the number of all possible dichotomous utility vectors. By \Cref{thm:dichotomous} we know that a worst-case instance exists among such instances. We start with a random starting point $q$ that represents instance  $(\prefp, \utp)$ and $\varepsilon=0.1$. In each iteration we define the set of all possible next steps as $S = \cup_{i \in [t]} \set{(q_1, \ldots, q_i+\varepsilon, \ldots, q_t), (q_1, \ldots, q_i-\varepsilon, \ldots, q_t)}$. We re-normalize all the vectors in $S$ to make them unit sum and compute their distortions using Linear Program \PP. Then  if there exists $q' \in S$ that represents instance  $(\prefp', \utp')$, where $\dist(f^*(\prefp'), \utp') > \dist(f^*(\prefp), \utp)$ we randomly jump to one of such vectors for the next iteration, and if non of them has a distortion greater than the distortion of $q$ then we decrease the value of $\varepsilon$ and repeat the procedure. We keep iterating while $\varepsilon \geq \tau_\varepsilon$ and report the distortion of instance optimal rule on the final instance as a lower bound on the distortion.

To make sure that we have a better bound we try different random starting points and report the maximum output.

While the algorithm does not return the global optimum (which would be the precise distortion of the instance-optimal rule), the optimal distortion it finds on any preference profile serves as a valid lower bound on the distortion of any rule.

\begin{table}[t]
\renewcommand{\arraystretch}{1.2}
    \centering
    \begin{tabular}{|c||c|c|c|c|}
    \toprule
    $\downarrow$ Rule, $m \rightarrow$&  2&  3&  4&  5 \\\midrule
         Best Randomized LB&  1.500&  1.835&  2.092&  2.303 \\ \midrule
         Best RPSR LB&  1.500&  1.870&  2.188&  2.473 \\
        Best RPSR UB & 1.500& 1.870& 2.188&2.474 \\\midrule
         Harmonic Rule&  1.556&  1.987&  2.354&  2.682 \\ 
         Randomized 2-Approval&  2.000&  2.414&  2.743&  3.078 \\  
         Randomized Borda&  1.522&  2.000&  2.556&  3.148 \\   
         Golden Rule&  1.714&  2.276&  2.751&  3.173 \\ 
          Randomized Plurality&  1.522&  2.155&  3.000&  4.000 \\ 
 Randomized 3-Approval& -& 3.000& 3.621&4.098 \\
         Randomized Veto&  1.522&  2.414&  3.621&  4.828 \\ \midrule
 Best PSR UB& 3& 5.5& 8.333&11.418 \\\bottomrule
    \end{tabular}
    \caption{The exact distortion of well-known RPSRs, and distortion bounds for the best RPSR, the best PSR, and the best randomized rule.}
    \label{tab:distortion_small_m}
\end{table}

Our results are summarized in \Cref{tab:distortion_small_m}. The benefits of using randomization are evident, as the distortions of deterministic PSRs are notably worse than those of even common RPSRs. On the other side of the spectrum, the best RPSRs are quite close (exactly matching for $m=2$) to the best randomized rules in general. We provide the best RPSRs achieving the displayed bounds in \Cref{tab:optimal_rules_2}. Among common RPSRs, most are near-optimal for $m=2$, but the differences between them are more pronounced for $m\geq 3$. Interestingly, with the exception of $m=2$, the Randomized Harmonic rule achieves the lowest distortion among all common RPSRs, even though the Golden Rule has a better asymptotic distortion (and the best asymptotic distortion among all RPSRs)~\cite{ebadian2023explainable}. This showcases that the hidden constants in the asymptotic bounds can make a difference for small values of $m$. 


\section{Experiments}
\label{sec:expr}


In this section, we present a comparative analysis between our linear program \PP and the LP of \citet{BCHL+15} (referred to as \btl) for computing the optimal distribution (and its corresponding distortion) on a given preference profile. We test on preference profiles generated synthetically as well as drawn from real-world datasets. To solve the LPs, we use Gurobi~\cite{gurobi}, with each run utilizing $4$ cores and $50\operatorname{GB}$ of memory.
\footnote{Code:  \url{https://github.com/latifian/Computational-Aspects-of-Distortion}}

\subsection{Synthetic data}
We generated instances from three statistical models: the \emph{Mallows} model with parameter $\phi \in \set{0.1, 0.2, 0.5, 0.8, 1}$, the \emph{Plackett-Luce} model, and random\footnote{We sampled the positions of agents and alternatives i.i.d. in $[0, 1]$ and derived preferences based on distances.} \emph{Single-Peaked} preferences; see~\cite{szufa2020drawing} for a description of these models. All instances consisted of $n = 1000$ agents. For $m \in \set{10, 20, 30, 40, 50, 60}$ alternatives and each model, we generated $10$ instances and calculated the average running time along with the standard error.

\begin{figure}[ht!]
    \centering
    \begin{minipage}[t!]{\textwidth}
        \centering
        \begin{subfigure}{0.45\textwidth}
            \centering
            \includegraphics[width=\textwidth]
            {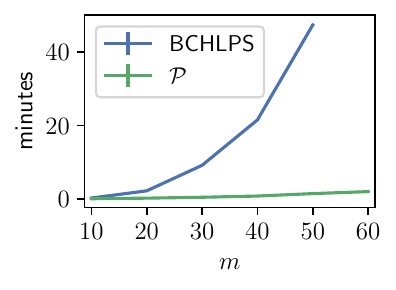}
            \caption{Mallows with $\phi=0.5$}
            \label{fig:mallows0.5}
        \end{subfigure}
        \begin{subfigure}{0.45\textwidth}
            \centering
            \includegraphics[width=\textwidth]{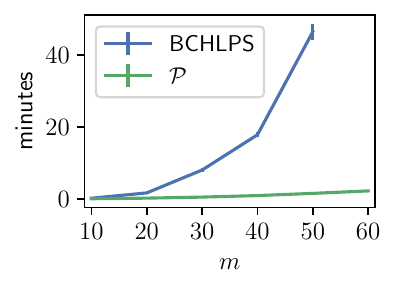}
            \caption{Plackett-Luce (P.L.)}
            \label{fig:placket2}
        \end{subfigure}
        \begin{subfigure}{0.45\textwidth}
            \centering
            \includegraphics[width=\textwidth]{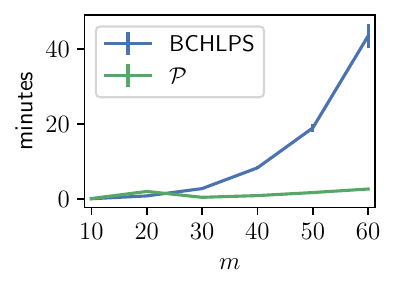}
            \caption{Single-Peaked (S.P.)}
            \label{fig:single-peaked2}
        \end{subfigure}
        \caption{Running time of \PP and \btl{} on synthetic data ($n=1000$).}
        \label{fig:synthetic-experiments2}
    \end{minipage}
    \label{fig:synthetic-all}
\end{figure}

\paragraph{Results}
\Cref{fig:synthetic-experiments2} and \Cref{tab:speedup} provide a summary of the running time comparison between the two LPs on synthetic data. Notably, \btl{} failed to terminate in 3 hours on every instance with $m=60$ under every model, except for the single-peaked model (\Cref{fig:single-peaked2}), whereas \PP completed in less than 150 seconds on each such instance. The speed-up (ratio of average running times) achieved by \PP is detailed in \Cref{tab:speedup}. For the larger values of $m \in \set{30, 40, 50}$, \PP consistently outperformed \btl{} across all statistical models, achieving $10\times$ to $30\times$ speedups, with the performance gap widening as $m$ increases. The only cases where \btl{} performed better were small instances ($m \in \set{10,20}$) generated using the Single-Peaked model. Our experiments demonstrate a significant improvement in the efficiency of \PP over \btl. 

\begin{table}[ht!]
\renewcommand{\arraystretch}{1.5}
        \centering
        \begin{tabular}{|c|c|c|c|c|c|}
        \hline
        $m$ & $\phi$=0.1 &$\phi$=0.5 &$\phi$=0.8 & S.P & P.L.  \\
        \hline
        10 & 3.74 & 2.07 & 3.1 & 0.77 & 3.29 \\ \hline
        20 & 9.71 & 5.13 & 8.08 & 0.43 & 7.28 \\ \hline
        30 & 19.6 & 9.09 & 15.79 & 5.48 & 15.42 \\ \hline
        40 & 26.8 & 23.37 & 20.18 & 8.54 & 18.91 \\ \hline
        50 & 32.28 & 31.06 & 28.89 & 10.61 & 29.84 \\
        \hline
        \end{tabular}
        \caption{Speed-up of \PP over \btl{}.}
        \label{tab:speedup}
    \end{table}

\subsection{PrefLib data}
We utilized the $7{,}742$ real-world preference profiles from PrefLib~\cite{mattei2013preflib} of type \emph{SOC} (strict order--complete order). The largest instances had as many as $n = 14{,}081$ voters and as many as $m=1080$ alternatives. 

\begin{figure}[ht!]
    \centering
    \begin{minipage}[t]{\textwidth}
        \centering
        \includegraphics[width=0.7\textwidth]{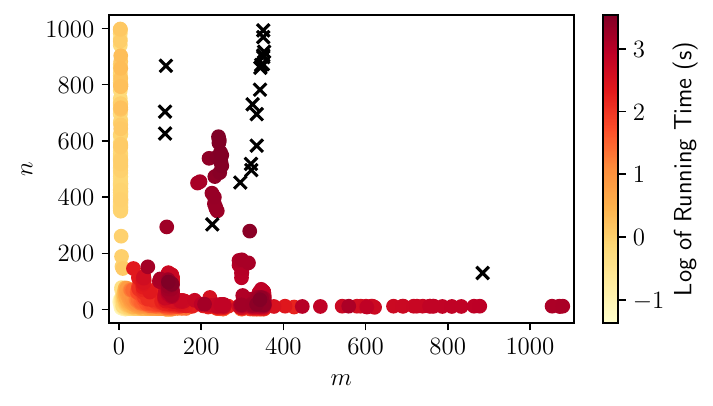}
        \caption{Performance of \btl{} on PrefLib data.}
        \label{fig:preflib-boutilier}
    \end{minipage}
    \begin{minipage}[t]{0.7\textwidth}
        \centering
        \includegraphics[width=\textwidth]{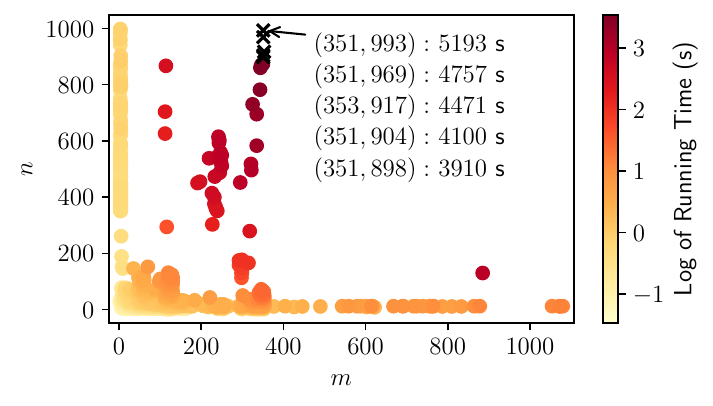}
        \caption{Performance of \PP on PrefLib data.}
        \label{fig:preflib-pp}
    \end{minipage}
    \label{fig:preflib}
\end{figure}

\paragraph{Results}
The running times of \btl{} and \PP on PrefLib data are illustrated in \Cref{fig:preflib-boutilier,fig:preflib-pp}, with color intensities representing $\log_{10}$ of the running time.

\Cref{tab:percentiles_performance} shows various percentiles of the running times of the two LPs across the PrefLib instances. 

\begin{table}[ht!]
\renewcommand{\arraystretch}{1.2}
\centering
\begin{tabular}{|c|c|c|c|c|c|c|}
\hline
Rule & $50\%$ & $90\%$ & $99\%$ & $99.5\%$ 
& mean \\
\hline
\btl{} & $51.63$s & $653.83$s& $1737.71$s& $2582.31$s 
& $218.28$s $^\dagger$ \\
\hline
\PP & $0.84$s & $4.6$s & $21.84$s &  $257.16$s 
& $9.42$s \\
\hline
\end{tabular}
\\
$\dagger$ excluding 20 instances where \btl{} did not finish in an hour.
\caption{Order statistics of running times on PrefLib.}
\label{tab:percentiles_performance}
\end{table}

Our LP \PP was successfully solved in under a minute in $99\%$ of the instances. On average across all instances, \PP completed in under $10$ seconds, marking a speed-up of more than $20\times$ over \btl. Additionally, instances denoted by $\times$ on \Cref{fig:preflib-boutilier,fig:preflib-pp} are the ones where the algorithms failed to terminate in an hour. These were 20 instances for \btl{} versus 5 instances for \PP; \PP completed these 5 instances within $1.5$ hours, while \btl{} still did not terminate on those 5 instances after 3 hours.



\section{Discussion and future work}
An interesting question is whether we can extend our results in \Cref{sec:small_vec} to larger values of $m$. The bottleneck is the running time of the quadratic program \QQ. Could there be a method that computes the distortion of a given RPSR faster? We conjecture that the associated computational problem is NP-hard, and therefore a polynomial-time algorithm should not be expected. Proving this conjecture is an interesting avenue for future work. Finally, could we extend our approach in \Cref{sec:qp-worst-case-instances} to voting rules beyond RPSRs? One candidate class of rules would be \emph{support voting schemes}, which, together with RPSRs, complete the class of truthful randomized rules on ordinal preferences~\cite{barbera1978nice,gibbard1977manipulation}. 

We studied single-winner selection, where the goal is to choose a single alternative as the winner. Some of our techniques may be applicable to related settings such as multiwinner selection, participatory budgeting, matching and assignment problems, incomplete preferences, domain restrictions, and metric distortion problems.

\section*{Acknowledgements}
Aris Filos-Ratsikas and Mohamad Latifian were supported by the UK Engineering and Physical Sciences Research Council (EPSRC) grant EP/Y003624/1.

\clearpage

\bibliographystyle{plainnat}
\bibliography{references}
\clearpage

\begin{appendices}
\section{The LP of \citet{BCHL+15}}
\label{app:boutilier-lp}

We include the LP of \Citet{BCHL+15} here for completeness. For details about the logic behind this linear program and its correctness, we refer the reader to the original paper. 

Let $\prefp$ be the queried preference profile, for which we want to find $\min_{p}\dist(p, \prefp)$. For each $\cd \in \cands$, define a variable $p_{\cd}$, and let $D = \left(\min_{p}\dist(p, \prefp)\right)^{-1}$ be a variable whose inverse corresponds to the optimal distortion. The LP of \citet{BCHL+15} is defined as follows.

\newcommand{\ineq}{\operatorname{INEQ}}
\begin{tcolorbox}
\centerline{\underline{Linear Program of \citet{BCHL+15}}}
\begin{align*}
    \text{max.} \quad & D & \\
    \text{s.t.} \quad 
    & \ineq(\{p_{\cd}\}_{\cd \in \cands}, D, \cd^*)~\text{is satisfied}  & \forall \cd^* \in \cands \\
    & \sum_{\cd \in \cands} p_{\cd} = 1 & \\
    & p_{\cd} \geq 0  & \forall \cd \in \cands,
\end{align*}

\vspace{2em}
$\ineq(\{p_{\cd}\}_{\cd \in \cands}, D, \cd^*):$
\vspace{0.5em}
\begin{align*}
& \sum_{i \in \voters} y_{i, \cd^*} \ge 0 \\
& D + \sum_{\cd \in \cands} x_{\cd, \cd^*} \le 0
\\
& p_{\cd} + x_{\cd, \cd^*} - y_{i, \cd^*} - z_{1, i, \cd^*}  \ge 0 \hspace{1.5cm} \forall i\in\voters, \cd = \pref_i(1)
\\
& p_{\cd} + x_{\cd, \cd^*} - y_{i, \cd^*} - z_{r, i, \cd^*}  + z_{r - 1, i, \cd^*} \ge 0
\\
\intertext{
\hspace{3.3cm}$
\forall i\in\voters, \cd = \pref_i(r) : r \in [2, m - 1] $
}
& p_{\cd} + x_{\cd, \cd^*} - y_{i, \cd^*} + z_{m - 1, i, \cd^*} \ge 0 \hspace{1cm} \forall i\in\voters, \cd = \pref_i(m)
\\
& x_{\cd, \cd^*} \ge 0 \hspace{4.5cm} \forall \cd \in \cands \setminus \{\cd^*\}
\\
& z_{r, i, \cd^*} \ge 0 \hspace{3.6cm} \forall r \in [m - 1], i \in \voters
\end{align*}
\end{tcolorbox}

For each $\cd^* \in \cands$, the number of variables and constraints in $\ineq(p, D, \cd^*)$ is $O(nm)$. Thus, the total number of variables and constraints is $O(nm^2)$.

\section{Comparing with the quadratic program of \cite{filos2014truthful}}\label{app:filos-ratsikas-miltersen}

As we mentioned in \Cref{sec:related}, \citet{filos2014truthful} devised a quadratic program for finding the distortion of a given RPSR, when the utilities of the agents are normalized according to the unit-range normalization.\footnote{\citet{filos2014truthful} only present their program in the context of $m=3$, but its definition is valid for any number of alternatives.} In this normalization, for every agent $i \in N$, we have that $\max_{a \in A} u_i(a)=1$, $\min_{a \in A} u_i(a)=0$, and $u_i(a) \in [0,1]$ for all $a \in A$. 

The quadratic programming formulation of \citep{filos2014truthful} is enabled by a structural characterization of the worst-case utility profiles, similar in spirit to that of our \Cref{thm:dichotomous}. In particular, they prove that the worst-case distortion is attained in what they refer to as ``quasi-combinatorial'' utility profiles, i.e., essentially utility profiles where the utility of an agent for any alternative is either $1$ or $0$. This enables them to restrict the set of possible utility vectors to a set of size $(m-1)\cdot m!$. Then, in a conceptually similar way to our approach in \Cref{sec:qp:worst-case-distortion-given-rpsr}, they define variables $x_i$ for the fraction of agents that have each of those utility vectors, and use those as variables of their optimization program. That program has linear constraints and a non-convex quadratic objective function, but the number of constraints is $(m-1)m!$. For $m=3$, in the unit-range case, that amounts to $12$ constraints, and \citeauthor{filos2014truthful} solve their quadratic program using active set methods, by enumerating over all possible subsets of active constraints.

For their approach to be even conceivably applicable to the unit-sum case, one would need first to prove a similar structural result for the worst-case utility profiles. Our \Cref{thm:dichotomous} clearly provides such a result. Still, for a given $m$, the number of possible $k$-dichotomous utility vectors is in fact $m\cdot m!$. While at first glance this might not sound much different from the $(m-1)m!$ used by \citeauthor{filos2014truthful} for unit-range, we remark that for $m=3$, this increases the number of variables (and constraints in their quadratic program) from 12 to 18. Solving their quadratic program with 18 constraints is clearly computationally infeasible, and hence their technique is inapplicable for unit-sum, even when $m=3$. 

Our crucial technical contribution here is the refinement of the worst-case characterization by means of \Cref{sw-order,eq-order}. In turn, this allows us to devise a (different) quadratic program with far fewer constraints, enabling us to handle up to $5$ alternatives in reasonable running time. 
\end{appendices}

\end{document}